\documentstyle{article}
\newcommand{\be}{\begin{equation}}
\newcommand{\bea}{\begin{eqnarray}}
\newcommand{\ba}{\begin{array}}
\newcommand{\bean}{\begin{eqnarray*}}
\newcommand{\ee}{\end{equation}}
\newcommand{\eea}{\end{eqnarray}}
\newcommand{\ea}{\end{array}}
\newcommand{\eean}{\end{eqnarray*}}
\newcommand{\hPi}{\mbox{$\widehat{\Pi}$}}

\newcommand{\bs}{\mbox{$\bar{s}$}}

\newcommand{\ups}{\mbox{$\upsilon$}}

\def \dsl {\partial \kern-.55em{/}}
\def \Dsl {D \kern-.65em{/}}
\def \qsl {q \kern-.45em{/}}
\def \slp {p \kern-.45em{/}}
\def \ksl {k \kern-.45em{/}}

\def \S {S-matrix~}

\def \se {self-energy~}
\def \ses {self-energies~}

\def \ind {independent~}
\def \ol {one-loop~}

\begin{document}
\begin{titlepage}
\rightline{ NYU-TH-96/02/02}
\rightline{February 1996}

\vspace*{1cm}
\Large
\begin{center}
{\bf  Consistency Condition for the Pinch Technique Self-Energies
 at Two Loops}
\end{center}

\vspace*{1cm}
\large
\begin{center}
{\bf Kostas Philippides and Alberto Sirlin }
\end{center}

\vspace*{1cm}
\large
\begin{center}
           {\it New York University, Department of Physics

      Andre and Bella Meyer Hall of Physics, 4 Washington Place

                    New York,NY 10003,USA.}
\end{center}

\vspace*{1cm}
\begin{center}
\large
{\bf ABSTRACT}
\end{center}
\vspace*{0.5cm}
\baselineskip=23pt

A simple and testable necessary condition for the gauge independence of the 
Pinch Technique self-energies at two loops is discussed. It is then shown 
that, in the case of the $Z$ and $W$ self-energies, the condition is indeed 
satisfied by the Papavassiliou-Pilaftsis formulation. 

\end{titlepage}

\large
\baselineskip=20pt

\setcounter{equation}{0}
 
The Pinch Technique (PT) is a convenient algorithm that automatically 
rearranges \S  elements of gauge theories into modified, gauge-independent 
self-energies, vertex, and box diagrams. In turn, the new corrections 
exhibit very desirable theoretical properties. For these reasons, the PT 
has been frequently employed in recent discussions of QCD and 
Electroweak Physics \cite{PT}. 
A temporary drawback is that the approach has been 
fully developed only at the one-loop level. Very recently, however, 
Papavassiliou and Pilaftsis (P-P) proposed a method to construct PT 
self energies at higher orders \cite{PaPi}. 

Calling $\hPi$ and $\Pi$ the PT and $R_{\xi}$ transverse self-energies, 
respectively, and focusing on the $Z$ case, one has 
\be 
 \hPi^{ZZ}(s) = \Pi^{ZZ}(s)+ (\Pi^{ZZ}(s))^P ~,
\label{Zse}
\ee
where the ``pinch part'' $(\Pi^{ZZ}(s))^P$ has the structure 
\be 
(\Pi^{ZZ}(s))^P = c_1(s-M^2_0)V^P(s)+c_2(s-M^2_0)^2B^P(s) - R^{ZZ}(s) ~.
\label{pinchpart}
\ee

In Eq.(\ref{pinchpart}) the bare mass $M_0$ is assumed to be defined 
in a gauge invariant manner, tadpole contributions are included in 
$\Pi^{ZZ}(s)$, $V^P(s)$ and $B^P(s)$ are the pinch parts from vertex and 
box diagrams, respectively, and $R^{ZZ}(s)$ is a residual amplitude of 
${\cal O}(g^4)$ proposed in Ref.\cite{PaPi}. It is discussed in detail 
later on at the ${\cal O}(g^4)$ level. Because of the limited knowledge 
currently available concerning multi-loop amplitudes in gauge theories, 
a general proof that $\hPi^{ZZ}(s)$ is gauge invariant in higher orders is 
not presently available. One of the aims of this report is to note that 
by judiciously restricting the domain of $s$ to lie in the neighborhood 
of $\bs$, the complex-valued position of the propagator's pole, one can 
obtain an expression for which the gauge independence 
 can be tested on the basis
of current knowledge. Specifically, we consider the neighborhood 
$|s-\bs| \le {\cal O}(g^2|\bs|)$, which roughly includes the resonance 
region. Recalling that $\bs-M_0^2 = {\cal O}(g^2)$, through ${\cal O}(g^4)$ 
Eqs.(\ref{Zse},\ref{pinchpart}) become 
\be
\hPi^{ZZ}(s) = \Pi_1^{ZZ}(s) + c_1(s-M^2_0)V_1^P(s) + \Pi_2^{ZZ}(\bs) 
- R_2^{ZZ}(\bs) + {\cal O}(g^6) ~, 
\label{Pi2}
\ee
where the indices $i=1,2$ denote  ${\cal O}(g^2)$ and ${\cal O}(g^4)$ 
contributions. Through ${\cal O}(g^4)$ the first two terms in the r.h.s. 
of Eq.(\ref{Pi2}) equal the \ol PT \se $\hPi_1^{ZZ}(s)$, which is $\xi-$\ind.
Its explicit expression \cite{DeSi2} is :
\be 
 \hPi_1^{ZZ}(s) = \Pi_1^{ZZ}(s)|_{\xi_i=1} - 4 g^2c_w^2(s-M^2_0)I_{WW}(s) ~,
\label{hPiol}
\ee
where $c_w^2$ is an abbreviation for $\cos\theta_w^2$, $\xi_i~(i=W,Z,\gamma)$ 
are the $R_{\xi}$ gauge parameters and 
\be
I_{ij}(q^2) = i\mu^{4-n}\int \frac{d^nk}{(2\pi)^n}
\frac{1}{\left(k^2-M_i^2\right)\left[(k+q)^2-M_j^2\right]} ~.
\label{Iij}
\ee
On the other hand, $\Pi_2^{ZZ}(\bs)$ in Eq.(\ref{hPiol}) is expected to be 
gauge dependent. Although this amplitude is not fully known,  
its gauge-dependent part can be isolated by a simple argument. Recalling 
that the pole position 
$\bs = M_0^2 + \Pi^{ZZ}(\bs) + [\Pi^{\gamma Z}(\bs)]^2/
[\bs-\Pi^{\gamma\gamma}(\bs)]$ is gauge invariant, through ${\cal O}(g^4)$ 
we have 
\be
\Pi_2^{ZZ}(\bs) = \bs - M_0^2 - \hPi_1^{ZZ}(\bs) + c_1(\bs- M_0^2)V_1^P(\bs) 
-[\Pi_1^{\gamma Z}(\bs)]^2/\bs ~,
\label{gdPi2}
\ee
where $\hPi_1^{ZZ}(\bs)$ is defined in Eq.(\ref{hPiol}). 
The amplitude $c_1V_1^P(s)$  
can be gleaned from Eqs.(12b,16d) of Ref.\cite{DeSi2} :
\be
c_1V_1^P(s) = -4g^2c_w^2\left[I_{WW}(s)+(\xi_W-1)\ups_W(s)/2\right] ~,
\label{c1V1}
\ee
where $\ups_W(s)$ is a $\xi_i$-dependent function defined in Eqs.(2-6) of 
Ref.\cite{DeSi1}. Recalling 
$\bs- M_0^2 = \Pi_1^{ZZ}(\bs)|_{\xi_i=1} + {\cal O}(g^4)$, we have 
\be 
c_1(\bs- M_0^2 )V_1^P(\bs) = -4g^2c_w^2\Pi_1^{ZZ}(\bs)|_{\xi_i=1}\left[
I_{WW}(\bs)+(\xi_W-1)\ups_W(\bs)/2\right] ~.
\label{olpinch}
\ee
On the other hand, using Eq.(8) of Ref.\cite{DeSi1},  
 one finds for general $\xi_i$ 
\bea
\frac{\left[\Pi_1^{\gamma Z}(\bs)\right]^2}{\bs} &=& 
\left[ \Pi_1^{\gamma Z}(\bs)|_{\xi_i=1} + g^2s_wc_w(\xi_W-1)\bs \ups_W(\bs)
\right]^2/\bs  + {\cal O}(g^6) \nonumber \\
&=& [\Pi_1^{\gamma Z}(\bs)|_{\xi_i=1}]^2/\bs + 2 g^2s_wc_w(\xi_W-1)\ups_W(\bs)
\Pi_1^{\gamma Z}(\bs)|_{\xi_i=1}\nonumber \\
&& + g^4s_w^2c_w^2(\xi_W-1)^2\bs \ups^2_W(\bs)  + {\cal O}(g^6) ~. 
\label{gZ2}
\eea
Combining Eqs.(\ref{gdPi2},\ref{olpinch},\ref{gZ2}) we obtain 
\bea
\Pi_2^{ZZ}(\bs) &=& \Pi_2^{ZZ}(\bs)|_{\xi_i=1} -2g^2(\xi_W-1)\ups_W(\bs)
\left[c_w^2\Pi_1^{ZZ}(\bs)|_{\xi_i=1} + c_ws_w 
\Pi_1^{\gamma Z}(\bs)|_{\xi_i=1}\right]  \nonumber \\
&& - g^4s^2_wc^2_w (\xi_W-1)^2 \bs \ups^2_W(\bs) ~. 
\label{GDPi2}
\eea

The terms proportional to $\xi_W-1$ and $(\xi_W-1)^2$ in Eq.(\ref{GDPi2}) 
represent the $\xi_i$-dependent parts of $\Pi_2^{ZZ}(\bs)$. It 
follows that, if the residual contributions $R_2^{ZZ}(\bs)$ are not 
included, Eq.(\ref{Pi2}) is gauge-dependent in ${\cal O}(g^4)$. 
Next we evaluate $R_2^{ZZ}(\bs)$. Following the P-P method  \cite{PaPi},
 $R_2$ 
is the contribution that must be added to the chain of $R_{\xi}$ 
transverse \ol \ses 
and corresponding pinch parts (Fig.1(b-d)), in order to convert it 
into the chain of \ol 
PT transverse \ses (Fig.1a). The explicit construction of   $R_2^{ZZ}(\bs)$ 
in the $\xi_i=1$ gauges has been given in Ref.\cite{PhiSi}. We must now 
 generalize this procedure to a general gauge. The chain of \ol PT \ses 
is by definition $\xi-$\ind and gives a contribution proportional to  
\be
\left[ \Pi_1^{ZZ}(s)|_{\xi_i=1} - (s-M_0^2)4g^2c_w^2I_{WW}\right]^2
/(s-M_0^2) + 
\left[ \Pi_1^{\gamma Z}(s)|_{\xi_i=1} - (2s-M_0^2)2g^2s_wc_wI_{WW}\right]^2/s
~,
\label{PTchain}
\ee
where we have employed Eqs.(16d,16b) of Ref.[3]. 
On the other hand, using the results of Refs.\cite{DeSi2}, \cite{DeSi1} 
and neglecting ${\cal O}(g^6)$, one finds that for 
$s=\bs$ and general $\xi_i$  the chain of $R_{\xi}$ \ol \ses and 
pinch parts contributes 
\bea
&\left[ \Pi_1^{ZZ}(\bs)|_{\xi_i=1} 
+ (\bs-M_0^2)2g^2c_w^2(\xi_W-1)\ups_W(\bs)\right]^2/(\bs-M_0^2)
\nonumber \\
&+ \left[ \Pi_1^{\gamma Z}(s)|_{\xi_i=1} 
+ \bs g^2s_wc_w(\xi_W-1)\ups_W(\bs)\right]^2
/\bs \hspace*{2cm}\nonumber \\
&-4g^2c_w^2\Pi_1^{ZZ}(\bs)|_{\xi_i=1}\left[I_{WW}(\bs)+(\xi_W -1)\ups_W(\bs)/2
\right] \hspace{1cm}~. 
\label{chains}
\eea
 The two first terms arise from the \se contributions 
in a general $R_{\xi}$ gauge [4]  through 
${\cal O}(g^4)$, while the third involves the contribution of 
\ses and pinch parts (Fig.1(b-d)) in the same approximation. 
Setting $s = \bs = M_0^2 + \Pi^{ZZ}_1(\bs) + ... $ in 
Eqs.(\ref{PTchain},\ref{chains}) and subtracting the two expressions we obtain 
\bea 
R_2^{ZZ}(\bs) = & -4g^2\left[c_w^2\Pi_1^{ZZ}(\bs) +s_wc_w \Pi_1^{\gamma Z}(\bs)
\right]|_{\xi_i=1} \left[ I_{WW}(\bs)+(\xi_W -1)\ups_W(\bs)/2 \right]
\nonumber \\
& + g^4s_w^2c_w^2\bs\left[4I^2_{WW}(\bs) - (\xi_W -1)^2\ups^2_W(\bs)\right]
+ {\cal O}(g^6) ~.
\label{R2}
\eea
Comparing Eq.(\ref{GDPi2}) with  Eq.(\ref{R2}) we see that the 
$\xi_W-$dependent contributions cancel in the combination 
$\Pi_2^{ZZ}(\bs)- R_2^{ZZ}(\bs)$. Thus, Eq.(\ref{Pi2}) is indeed 
gauge-\ind through  ${\cal O}(g^4)$ if $R_2^{ZZ}(\bs)$ is evaluated according 
to the P-P method. In the neighborhood  $|s-\bs| \le {\cal O}(g^2|\bs|)$,
 Eq.(\ref{Pi2}) becomes 
\bea  
\hPi^{ZZ}(s) = & \Pi_1^{ZZ}(s)|_{\xi_i=1}+\Pi_2^{ZZ}(\bs)|_{\xi_i=1} 
-4g^2c_w^2(s-\bs)I_{WW}(\bs) \nonumber \\
&+ 4g^2s_wc_w\Pi_1^{\gamma Z}(\bs)|_{\xi_i=1}I_{WW}(\bs) 
-4g^4s^2_wc_w^2 \bs I^2_{WW}(\bs) + {\cal O}(g^6) ~.
\label{aroundpole}
\eea
For $s = \bs$, Eq.(\ref{aroundpole}) reduces to Eq.(3.16) of Ref.\cite{PhiSi}.
Using Eq.(\ref{aroundpole}) one finds  that in the PT approach 
the denominator in the $Z$ propagator can be written as 
\be
s-M_0^2-\hPi^{ZZ}(s) - [\hPi^{\gamma Z}(s)]^2/s = 
(s-\bs)\left[1-\frac{d}{ds}\hPi_1^{ZZ}(s)|_{s=\bs}\right] + {\cal O}(g^6) ~,
\label{denomZ}
\ee
where it is understood that $|s-\bs| \le {\cal O}(g^2|\bs|)$. 
In order to derive Eq.(15), it is convenient to add and subtract 
$\bs$ in the l.h.s., employ 
$\bs - M_0^2 = \Pi^{ZZ}(\bs) + (\Pi^{\gamma Z}(\bs))^2/\bs$ + ... , 
and recall the expression for $\hPi^{\gamma Z}$ given in 
Eq.(16b) of Ref[3]. 
Eq.(\ref{denomZ}) explicitly shows two important properties: 1) as it 
involves the PT \se $\hPi_1^{ZZ}(s)$, it is manifestedly $\xi_i-$\ind 
2) the zero of Eq.(\ref{denomZ}) occurs at  $s = \bs$, so that the 
pole position is not displaced. For the $Z$ case, the latter
 property was already 
derived in the particular case of the $\xi_i =1$ gauges \cite{PhiSi}. 
As explained in Refs.\cite{PhiSi}, \cite{SiMass}, 
using the scaling approximation 
for ${\cal I}m\hPi_1^{ZZ}(s)$ one can transform  Eq.(\ref{denomZ}) into 
the characteristic $s-$dependent Breit-Wigner resonance employed in the LEP 
analysis, so that the connection with the LEP observables becomes explicit.

One can readily carry out the same analysis for the $W$ \se. In this case 
there are no mixing complications but the $R_{\xi}$ gauge dependence is 
governed by three parameters $\xi_i ~(i=W,Z,\gamma)$. One finds 
\be
\Pi_2^{WW}(\bs) = \Pi_2^{WW}(\bs)|_{\xi_i=1} -g^2\Pi_1^{WW}(\bs)|_{\xi_i=1}
F(\xi_i,\bs) + {\cal O}(g^6) ~,
\label{gdWPi}
\ee
\be
F(\xi_i,\bs) = c^2_w[(\xi_W-1)\ups_{WZ}(\bs) + ( W\leftrightarrow Z)]
+ s^2_w[(\xi_W-1)\ups_{W\gamma}(\bs)+ ( W \leftrightarrow\gamma)] ~,
\label{F}
\ee
where $(i\leftrightarrow j)$ is obtained from the preceding term 
by interchanging the 
indices in $(\xi_i-1)\ups_{ij}(\bs)$. The gauge dependence is contained 
in $F(\xi_i,\bs)$. Following the P-P method we obtain 
\be
R_2^{WW}(\bs) = -4g^2\Pi_1^{WW}(\bs)|_{\xi_i=1}\left[ 
c_w^2I_{ZW}(\bs)+s^2_wI_{\gamma W}(\bs)\right]
-g^2\Pi_1^{WW}(\bs)|_{\xi_i=1}F(\xi_i,\bs) ~.
\label{R2W}
\ee
Again $\Pi_2^{WW}(\bs)-R_2^{WW}(\bs)$ is $\xi_i-$\ind and in the interval 
$|s-\bs| \le {\cal O}(g^2|\bs|)$ we find 
\be
\hPi^{WW}(s) = \Pi_1^{WW}(s)|_{\xi_i=1} + \Pi_2^{WW}(\bs)|_{\xi_i=1} 
-4g^2(s-\bs)\left[ c_w^2I_{ZW}(\bs)+s^2_wI_{\gamma W}(\bs)\right]
+ {\cal O}(g^6) ~.
\label{poleWPi}
\ee
Alternatively, the $\xi_i-$independence of $\hPi_2^{WW}(\bs)$ can be derived 
by directly evaluating 
$\hPi_1^{WW}(\bs)+\hPi_2^{WW}(\bs)-\Pi_1^{WW}(\bs)-\Pi_2^{WW}(\bs)$ 
in a general $\xi_i$ gauge \cite{PaPi}. 
Using Eq.(\ref{poleWPi}) the propagator's denominator becomes 
\be
s-(M_0^W)^2 - \hPi^{WW}(s) = (s-\bs)\left[1-\frac{d}{ds}\hPi_1^{WW}(s)|_{s=\bs}
\right] + {\cal O}(g^6) ~,
\label{denomW}
\ee
in analogy with the $Z$ case. 

In summary, by restricting $s$ to the neighborhood 
$|s-\bs| \le {\cal O}(g^2|\bs|)$ of the propagator's pole one can test the 
 gauge dependence of 
the $ZZ$ and $WW$ \ses through ${\cal O}(g^4)$. In both cases we 
find that the P-P method leads to gauge \ind amplitudes. 
Because of our restriction to the resonance region, 
and our neglect of ${\cal O}(g^6)$ terms,  this test amounts 
to a necessary rather than a sufficient condition. 
On the other hand, it is important to emphasize that this domain is of 
special physical significance. 

\vspace*{1cm}

This work was supported in part by the National Science Foundation 
under grant No. PHY-9313781. 

\vspace*{1.5cm}

\Large
\noindent{\bf Figure Caption}

\vspace*{.5cm}
\large
Chain of one-loop transverse PT self-energies through ${\cal O}(g^4)$ 
and a class of related pinch parts.

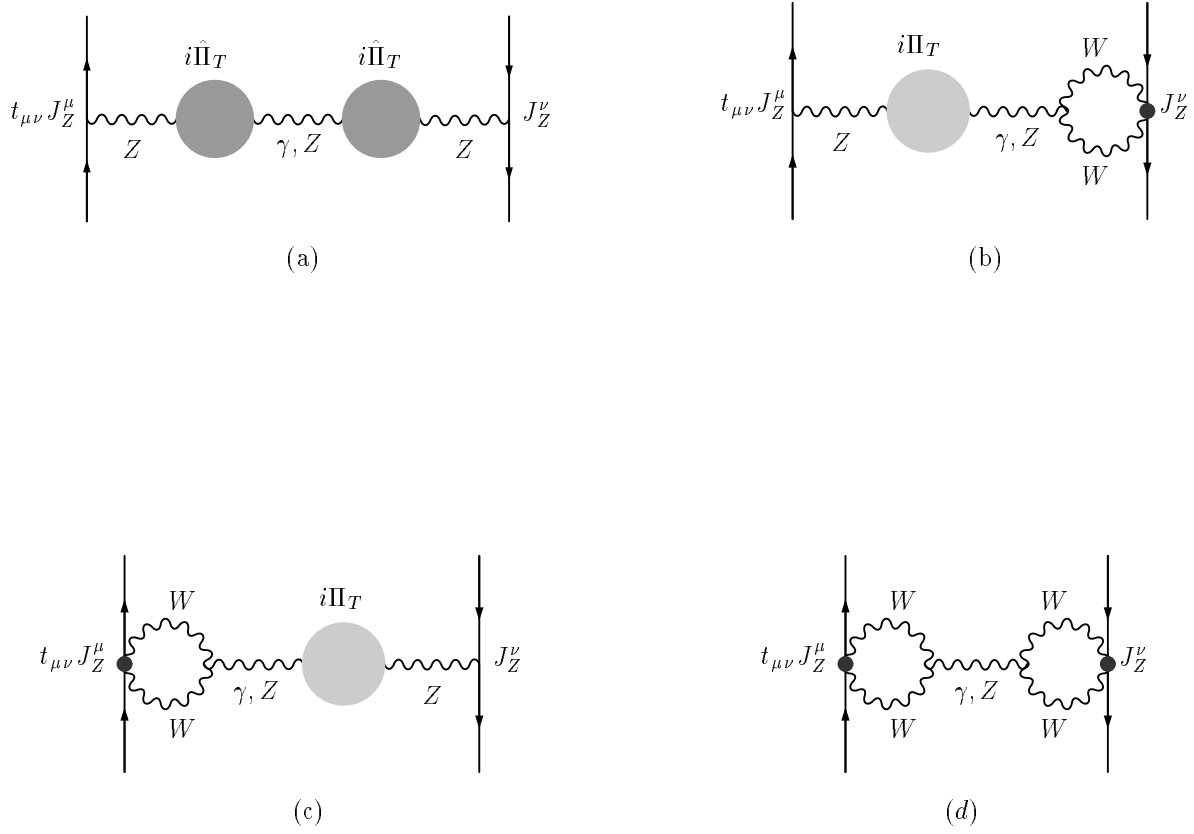
\end{document}